\documentclass[12pt]{article}
\usepackage{graphicx}

\begin{document}

\newcommand{\be}{\begin{eqnarray}}
\newcommand{\ee}{\end{eqnarray}}
\newcommand{\ben}{\begin{eqnarray*}}
\newcommand{\een}{\end{eqnarray*}}
\newcommand{\la}{\langle}
\newcommand{\ra}{\rangle}
\def\Sp{\mathop{\rm Sp}\nolimits}

\title{Casimir effect in deformed field}
\author{I.~O.~Vakarchuk\\
{\em Department for Theoretical Physics,}\\
{\em Ivan Franko National University of Lviv}\\
{\em 12 Drahomanov St., Lviv, UA--79005, Ukraine}
}
\maketitle

\abstract{
The Casimir energy is calculated in one-,
two-, and three-dimensional spaces for the field with
generalized coordinates and momenta satisfying the deformed
Poisson brackets leading to the minimal length.


}

\section{Initial equations}

The Casimir effect, starting from the seminal work \cite{1}, has
been studied in the course of several decades for various physical systems
from the problems of solid state physics to those of cosmology.
Recently, some interest arose in the study of this effect in the
so-called deformed spaces, i.\,e., in the spaces with deformed Poisson
brackets and, in particular, in those leading to the minimal
lengths \cite{2,3,4}.

In our work, we will study the Casimir effect for the deformed
electro-magnetic field with the Hamiltonian
\be
\hat H=\sum_{\bf k}\sum_\alpha\left({\hat P_{{\bf k},\alpha}^2\over 2}
+{\omega_k^2 \hat Q_{{\bf k},\alpha}^2\over 2}\right),
\ee
where ${\bf k}$ is the wave vector, $\alpha$ is the polarization index, the frequency $\omega_k=ck$,
$c$ is the speed of light in a vacuum, $\hat Q_{{\bf k},\alpha}$, $\hat P_{{\bf k},\alpha}$ are the operators
of generalized coordinates and momenta that satisfy the deformed Poisson brackets:
\be
\hat Q_{{\bf k},\alpha} \hat P_{{\bf k},\alpha}- \hat P_{{\bf
k},\alpha}\hat Q_{{\bf k},\alpha}= i\hbar \Big(1+\beta \hat P_{{\bf
k},\alpha}^2\Big),
\label{1.2}
\ee
$\beta\geq0$ is the deformation parameter, all the other commutators equal zero.
Let us also note
that the deformation parameter can depend on ${\bf k}$ and
$\alpha$. In this work,  $\beta$ is put constant.

As is known, such commutation relations lead to the existence
of the minimal length $\sqrt{\la\hat Q_{{\bf k},\alpha}\ra^2}=\hbar\sqrt \beta$
in the space of field coordinates \cite{5}. Obviously, in this case one has
not the deformation of a real space but that of the field itself.
Proceeding to new operators $\hat q_{{\bf k},\alpha}$, $\hat p_{{\bf k},\alpha}$,
\be
\hat q_{{\bf k},\alpha}=\hat Q_{{\bf k},\alpha},\ \ \ \ \
\hat P_{{\bf k},\alpha}={1\over \sqrt \beta}\tan (\hat p_{{\bf
k},\alpha}\sqrt \beta),
\ee
it is easy to show that they are canonically conjugated,
\be
\hat q_{{\bf k},\alpha}\hat p_{{\bf k},\alpha}-\hat p_{{\bf k},\alpha}\hat q_{{\bf
k},\alpha}=i\hbar,
\ee
and the Hamiltonian equals
\be
\hat H =\sum\limits_{\bf k}\sum\limits_\alpha
\left({\omega_k^2 \hat q_{{\bf k},\alpha}^2\over 2}
+{\tan^2(\hat p_{{\bf k},\alpha}\sqrt \beta)\over 2\beta}\right).
\label{1.5}
\ee

The equations of motion of the field were studied in \cite{6}, where the Hamiltonian
was presented in the form of the expansion over the powers of
ordinary creation and annihilation operators.
Thus, the field equations are non-linear and, generally speaking,
might be analyzed, as a rule, by means of the perturbation theory.
A non-linear field described by $q$-oscillators was studied in \cite{7}.

The energy levels of the harmonic oscillator Hamiltonian (\ref{1.5}) with the commutation
relations (\ref{1.2}) are well known \cite{5,8}.
Therefore, the energy levels of the deformed field are as follows:
\be
E_{\ldots,N_{{\bf k},\alpha},\ldots}&=&\sum_{\bf k}\sum_\alpha \hbar
\omega_k\Bigg[\left(N_{{\bf k},\alpha}+{1\over 2}\right)
\sqrt{1+\left({\beta \hbar \omega_k\over 2}\right)^2}\nonumber
\\
&+&{\beta \hbar \omega_k\over 2}
\left(N_{{\bf k},\alpha}^2+N_{{\bf k},\alpha}+{1\over
2}\right)\Bigg],
\ee
where the quantum numbers  $N_{{\bf k},\alpha}=0,1,2,\ldots\;$.

The energy of the vacuum state of the field, when $N_{{\bf k},\alpha}=0$,
equals
\be
E_{\ldots,0,\ldots}&=&\sum_{\bf k}\sum_\alpha {\hbar
\omega_k\over 2}\left(\sqrt{1+\left({\beta \hbar \omega_k\over 2}\right)^2}
+{\beta \hbar \omega_k\over 2}\right).
\label{1.4}
\ee
The aim of the present work is the calculation of the Casimir
energy for the deformed space as a function of the deformation
parameter $\beta$ proceeding from Eq.~(\ref{1.4}).

\section{Casimir energy in a one-dimensional space}

Let the scalar field be concentrated in a one-dimensional (1D) space on
the segment of length $a$ along the $x$ axis between the points $x=0$ and $x=a$,
in which it equals zero. At such conditions, the wave vector in
Eq.~(\ref{1.4}) $k={\pi n/a}$, $n=1,2,\ldots$ and there is no polarization.
By definition, the Casimir energy $\varepsilon$ equals the
difference of the energy density (\ref{1.4}) of a vacuum in volume $a$ and that in an infinite volume:
\be
\varepsilon&=&{1\over a^2} \sum_{n=1}^\infty
{\hbar c \pi n\over 2} \left(\sqrt{1+\left({\beta \hbar c\over
2a}n\right)^2}+{\beta \hbar c\over 2a}n\right)
\nonumber\\
&-&{1\over 2\pi}\int\limits_{-\infty}^\infty {\hbar c |k|\over 2} \left(\sqrt{1+\left({\beta \hbar c\over
2}k\right)^2}+{\beta \hbar c\over 2}\,|k|\right)\, dk.
\ee
To regularize this expression one can introduce a cut-off function $e^{-\nu k}$, $\nu>0$,
which ensures both the convergence of the summation over $n$ and the integration
over $k$. Further we make the change of the variable of integration $k=\pi n/a$ and get as a result:
\be
\varepsilon&\mathop{=}\limits_{\nu\to 0}&{\hbar c \pi\over 2a^2}
\Bigg[ \sum_{n=1}^\infty e^{-\nu n} n  \left(\sqrt{1+\left(\beta^{*}n\right)^2}+\beta^{*}n\right)
\nonumber\\
&-&\int\limits_{0}^\infty e^{-\nu n}n
\left(\sqrt{1+\left(\beta^{*}n\right)^2}+\beta^{*}n\right)\,
dn\Bigg],
\label{nn}
\ee
where the dimensionless deformation parameter is introduced
$$
\beta^{*}=\beta{\hbar c \pi\over 2a}.
$$

Let us now apply the Abel--Plana formula \cite{9} to calculate the
sum over $n$ in Eq.~(\ref{nn}):
\be
\sum_{n=1}^\infty f(n)=\int\limits_0^\infty f(n)\, dn-{f(0)\over 2}
+i\int\limits_0^\infty {f(it)-f(-it)\over e^{2\pi t}-1}\, dt.
\label{v2.10}
\ee
We have
\be
f(n)=e^{-\nu n}n(\sqrt{1+(\beta^{*}n)^2}+\beta^{*}n).
\ee
Now, taking into account (\ref{v2.10}), the integrals over $n$ in Eq.~(\ref{nn}) cancel out, $f(0)=0$,
and the integration over $t$ (with the branching of the integrand at $t=1/\beta^{*}$ taken into
consideration) gives:
\be
\varepsilon=-{\hbar c\pi\over a^2} \int\limits_0^{1/\beta^{*}}
{t\sqrt{1-(\beta^{*}t)^2}\over e^{2\pi t}-1}\, dt.
\label{ee}
\ee
The value $\beta^{*}=0$ leads to the known result \cite{8,10}:
\be
\varepsilon=-{\hbar c\pi\over a^2} \int\limits_0^\infty {t\over
e^{2\pi t}-1}\, dt =-{\hbar c \pi\over 24 a^2}.
\ee

Let us now find the expansion of $\varepsilon$ at $\beta^{*}>1$.
For this purpose one can write expression (\ref{ee}) as follows:
\be
\varepsilon=-{\hbar c\over 2\beta^{*}a^2} \int\limits_0^1
\sqrt{1-x^2}{2\pi x/\beta^{*}\over e^{2\pi x/\beta^{*}}-1}\, dx.
\ee
Let us expand the second factor under the integral over the powers of  $2\pi x/\beta^{*}$
using the definition of the Bernoulli numbers $B_n$ \cite{11,12}:
\be
\varepsilon=-{\hbar c\over 2\beta^{*}a^2}\sum_{n=0}^\infty
{B_n\over n!}\left({2\pi\over \beta^{*}}\right)^n \int\limits_0^1 x^n
\sqrt{1-x^2}\, dx.
\ee
This integral is Euler's B-function, and the Bernoulli numbers are expressed via
Riemann's $\zeta$-function \cite{11,12}. As a result we obtain:
\be
\varepsilon&=&
{\hbar c \pi\over 2a^2} \Bigg(\sum_{n=1}^\infty
(-)^n {\Gamma (n+1/2)\zeta (2n)\over (n+1)!2\sqrt \pi
\beta^{*2n+1}}-{1\over 4\beta^{*}}+{1\over
3\beta^{*2}}\Bigg)
\nonumber \\
&=&
-{\hbar c\pi\over 24 a^2} \left(
{3\over \beta^{*}}-{4\over \beta^{*2}}+{\pi^2\over 4\beta^{*3}}-{\pi^4\over 120\beta^{*5}}
+{\pi^6\over 2016\beta^{*7}}+\ldots
\right).
\label{bb}
\ee

It is interesting to find this result directly from Eq.~(\ref{nn}).
For this purpose, let us single out the asymptotics $2\beta^{*}n^2+1/2\beta^{*}$, $n\to \infty$,
of the expression in parentheses in formula (\ref{nn}) and write
it as follows:
\be
\varepsilon&\mathop{=}\limits_{\nu\to 0}&{\hbar c \pi\over 2a^2}
\Bigg\{\sum_{n=1}^\infty e^{-\nu n}   \left[
\beta^{*}n^2 \left(\sqrt{1+{1\over(\beta^{*}n)^2}}-1\right)
-{1\over 2\beta^{*}}\right]
\nonumber\\
&-&\int\limits_{0}^\infty e^{-\nu n}
 \left[\beta^{*}n^2\left(\sqrt{1+{1\over(\beta^{*}n)^2}}-1\right)
-{1\over 2\beta^{*}}\right]\,dn\nonumber\\
&+&\sum_{n=1}^\infty e^{-\nu n}
\left(2\beta^{*}n^2+{1\over 2\beta^{*}}\right)
- \int\limits_{0}^\infty e^{-\nu n} \left(2\beta^{*}n^2+{1\over
2\beta^{*}}\right)\, dn
\Bigg\}.
\label{vv}
\ee
At $\nu \to 0$ the first integral in this formula equals  $(-1/3\beta^{*2})$,
and the difference of two last expressions equals $(-1/4\beta^{*})$.
Finally we get
\be
\varepsilon={\hbar c \pi\over 2a^2}
\Bigg\{\sum_{n=1}^\infty \left[
\beta^{*}n^2 \left(\sqrt{1+{1\over(\beta^{*}n)^2}}-1\right)
-{1\over 2\beta^{*}}\right]-{1\over 4\beta^{*}}+{1\over
3\beta^{*2}}\Bigg\}.
\label{v2.13}
\ee
As the sum over $n$ converges, the cut-off parameter $\nu$ can be
put zero.
Proceeding further with a formal expansion of the square root in
the series of $1/\beta^{*2}$ and using the definition of Riemann's
$\zeta$-function, we come to (\ref{bb}).

\begin{figure}[h]
\centerline{\includegraphics[width=0.8\textwidth,scale=0.6,clip]{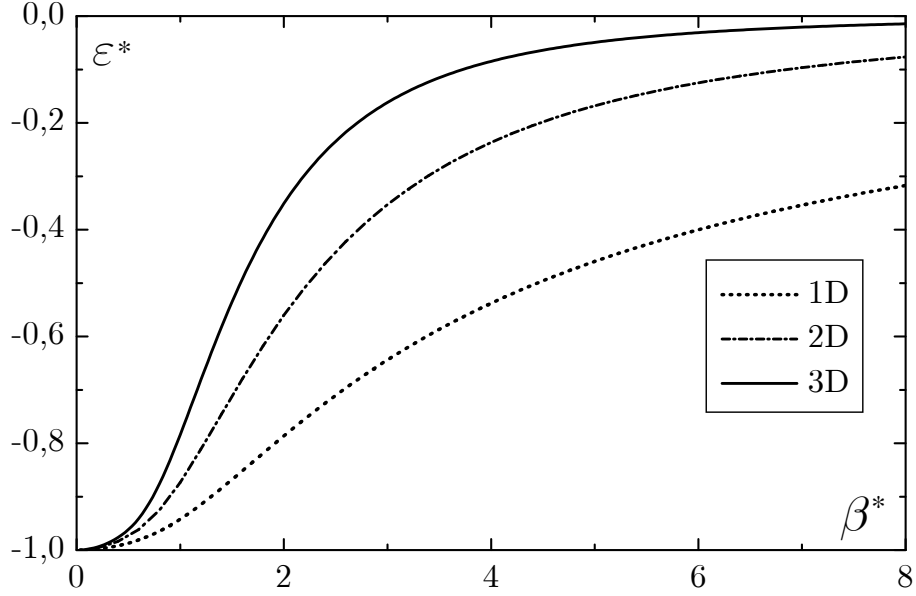}}
\medskip
\caption{Casimir energy as a function of the deformation parameter.}
\end{figure}

To make the computer calculation more convenient, formula
(\ref{v2.13}) can be transformed to the following form:

\be
\varepsilon=-{\hbar c \pi\over 2\beta^{*}a^2}\left[{1\over 2} \sum_{n=1}^\infty
{1\over \left(\sqrt{1+(\beta^{*}n)^2}
+\beta^{*}n\right)^2}+{1\over 4}
-{1\over 3\beta^{*}}\right].
\label{v2.19}
\ee
It should be noted, however, that the calculation of $\varepsilon$ using (\ref{ee})
is more effective in comparison with the summation in (\ref{v2.19}). This is the consequence
of a fast convergence of the integral in (\ref{ee}).

Let us now make the expansion of the Casimir energy at small
values of the deformation parameter.
One can apply the Euler--Maclaurin formula directly to
(\ref{nn}), as Casimir did in \cite{1}.
However, the result can be obtained in a shorter way if expression
(\ref{nn})is written as follows:
\be
\varepsilon&\mathop{=}\limits_{\nu\to 0}&-{\hbar c \pi\over 2a^2}
{d\over d\nu}\sqrt{1+\left(\beta^{*}{d\over d\nu}\right)^2}
\left(\sum_{n=1}^\infty e^{-\nu n}-\int\limits_0^\infty e^{-\nu n}\,
dn\right)\nonumber\\
&\mathop{=}\limits_{\nu\to 0}&-{\hbar c \pi\over 2a^2}
{d\over d\nu}\sqrt{1+\left(\beta^{*}{d\over d\nu}\right)^2}
\left({1\over e^\nu-1}-{1\over \nu}\right)\\
&\mathop{=}\limits_{\nu\to 0}&-{\hbar c \pi\over 2a^2}
{d\over d\nu}\sqrt{1+\left(\beta^{*}{d\over d\nu}\right)^2}
\sum_{n=1}^\infty {B_n\over n\,!}\nu^{n-1}.
\nonumber
\label{mm}
\ee
Here we again used the definition of the Bernoulli numbers $B_n$ \cite{11,12}.
Let us now make a formal expansion of the square root over the
powers of $\beta^{*2}$, take the derivatives over $\nu$ at $\nu\to 0$
and after simple transformations find the following asymptotic
series:
\be
\varepsilon&=&-{\hbar c \pi\over 2a^2}
\left({1\over 12}+\sum_{n=0}^\infty (-)^n{\Gamma(n-1/2)\over (n+1)!\,4\sqrt\pi}
B_{2n+2}\beta^{*2n}\right)\nonumber
\\
&=&-{\hbar c \pi\over 24a^2}
\left(1-{\beta^{*2}\over 20}-{\beta^{*4}\over 168}-{\beta^{*6}\over 320}-{5\over 1408}
\beta^{*8}-\ldots\right).
\ee

The results of the calculation of the Casimir energy
(\ref{ee}), (\ref{v2.19})
\ben
\varepsilon_{\rm 1D}^{*}=\varepsilon\left/{\hbar c\pi\over 24
a^2}\right.
\een
are presented in Fig.\,1 (curve 1D).
As one can see,  the Casimir energy for all the values of the
parameter $\beta^{*}$ is larger than that of the undeformed field
and tends to zero at $\beta^{*}\to \infty$. Therefore, the
deformation leads to the decrease of the attraction of the domain
boundaries localizing the field caused by the polarization of a
vacuum.

\section{A three-dimensional case}

Let us now calculate the Casimir energy for the field concentrated
is a three-dimensional (3D) space between two infinite parallel plates
separated by the distance $a$. The field equals zero on the plates.
The wave vector $k=\sqrt{k_x^2+k_y^2+k_z^2}$,
$-\infty<k_x<\infty$, $-\infty<k_y<\infty$, $k_z=\pi n/a$,
$n=0,1,2,\ldots\,$, and the sum
$$
\sum_{\bf k}\ldots \to {1\over (2\pi)^2}\int_{-\infty}^\infty dk_x
\int_{-\infty}^\infty dk_y \sum_{n=0}^\infty \ldots\;.
$$
Unlike the 1D case, in three dimensions the term with $n=0$ must
be taken into account in the expression for the ground state
energy (\ref{1.4}). However, the polarization index $\alpha$ in
this case has only one value which is a consequence of the field
transversality.
Therefore, by definition the Casimir energy equals
\be
\varepsilon&=&{1\over a}{1\over (2\pi)^2}\int\limits_{-\infty}^\infty
dk_x\int\limits_{-\infty}^\infty dk_y{\hbar c\over 2}\sqrt
{k_x^2+k_y^2}\nonumber\\
&\times&\left[\sqrt{1+\left({\beta\hbar c\over 2}\right)^2(k_x^2+k_y^2)}
+{\beta\hbar c\over 2}\sqrt {k_x^2+k_y^2}
\right]\nonumber\\
&+&{1\over a}\sum_\alpha{1\over (2\pi)^2}\int\limits_{-\infty}^\infty
dk_x\int\limits_{-\infty}^\infty dk_y \sum_{n=1}^\infty {\hbar c\over 2}\sqrt
{k_x^2+k_y^2+\left({\pi\over a}n\right)^2}\nonumber\\
&\times&
\left[\sqrt{1+\left({\beta\hbar c\over 2}\right)^2
\left(k_x^2+k_y^2+\left({\pi\over a}n\right)^2\right)}
+{\beta\hbar c\over 2}\sqrt{k_x^2+k_y^2+
\left({\pi\over a}n\right)^2}\right]\label{v3.20}\\
&-&\sum_\alpha {1\over (2\pi)^3}\int\limits_{-\infty}^\infty
dk_x\int\limits_{-\infty}^\infty dk_y \int\limits_{-\infty}^\infty dk_z
{\hbar c\over 2}\sqrt
{k_x^2+k_y^2+k_z^2}\nonumber\\
&\times& \left[\sqrt{1+\left({\beta\hbar c\over2}\right)^2
\left(k_x^2+k_y^2+k_z^2\right)}+{\beta\hbar c\over 2}\sqrt{k_x^2+k_y^2+k_z^2}\right]\nonumber
\ee

The integration is made in polar coordinates, $q=\sqrt{k_x^2+k_y^2}$,
with a further change of variables
$k=\sqrt{q^2+(\pi n/a)^2}$, replacing $k$ with $q$,
and introducing the cut-off function $e^{-\nu q}$. The result of these transformations is as follows:
\be
\varepsilon&\mathop{=}\limits_{\nu \to 0}&{\hbar c\over 2\pi a}\left\{{1\over 2} \int\limits_0^\infty dq\,
q^2 e^{-\nu q} \left[\sqrt{1+\left({\beta\hbar cq\over 2}\right)^2 }+{\beta \hbar cq\over
2}\right]\right.\nonumber\\
&+&\sum_{n=1}^\infty \int\limits_{\pi n/2}^\infty dq\,
q^2 e^{-\nu q}\left. \left[\sqrt{1+\left({\beta\hbar cq\over 2}\right)^2 }+{\beta \hbar cq\over
2}\right] \right\}\nonumber\\
&-&{\hbar c\over 2\pi^2}\int\limits_0^\infty dk_z \int\limits_{k_z}^\infty dq\,
q^2 e^{-\nu q} \left[\sqrt{1+\left({\beta\hbar cq\over 2}\right)^2}+{\beta \hbar cq\over
2}\right].
\label{jjj}
\ee
The Abel--Plana formula (\ref{v2.10}) with the function
\be
f(n)=\int\limits_{\pi n/2}^\infty q^2 e^{-\nu q}\left[\sqrt
{1+\left({\beta \hbar c q\over 2}\right)^2}+{\beta \hbar c q\over2}
\right]\, dq.
\ee
is applied to the second term in braces. It is now seen that the
term with $f(0)$ from Eq.~(\ref{v2.10}) cancels the first term in (\ref{jjj}),
and the last term in (\ref{jjj}) cancels the first integral from
the Abel--Plana formula (\ref{v2.10}). After a simple transformation we obtain:
\be
\varepsilon=-{\hbar c \pi^2\over 2a^4}\int\limits_0^{1/\beta^{*}}
dt {1\over e^{2\pi t}-1}\int\limits_{-t}^t x^2\sqrt{1-(\beta^{*}x)^2}\,
dx.
\ee
Taking the integral over $x$ we finally find:
\be
\varepsilon=-{\hbar c \pi^2\over 8\beta^{*2}
a^4}\int\limits_0^{1/\beta^{*}}{(2\beta^{*2}t^3-t)\sqrt{1-(\beta^{*}t)^2}+{\rm
arcsin}(\beta^{*}t)/\beta^{*}\over e^{2\pi t}-1}\,dt.
\label{v3.24}
\ee

At $\beta^{*}\to 0$ the numerator of the integrand in
(\ref{v3.24}) tends to $8\beta^{*2}t^3/3$ and
\be
\varepsilon=-{\hbar c\pi^2\over 3a^4} \int\limits_0^\infty
{t^3\over e^{2\pi t}-1}\, dt=-{\hbar c \pi^2\over 720 a^4}.
\ee
This is Casimir's result \cite{1}.

Let us now find the expansion of $\varepsilon$ over the powers of $1/\beta^{*}$
at $\beta^{*}\neq 0$. Expression (\ref{v3.24}) after the change of variable $x=\beta^{*}t$ is written as follows:
\be
\varepsilon=-{\hbar c \pi\over 16 a^4\beta^{*3}}\int\limits_0^1
\left[(2x^2-1)\sqrt{1-x^2}+{{\rm arcsin} x\over x}\right]{2\pi
x/\beta^{*}\over e^{2\pi x/\beta^{*}}-1}\, dx.
\ee
Using the expansion of the second factor under the integral over
the powers of $2\pi x/\beta^{*}$ \cite{11,12} we find:
\be
\varepsilon=-{\hbar c \pi\over 16a^4 \beta^{*3}}\sum_{n=0}^\infty
{B_n\over n!}\left({2\pi\over \beta^{*}}\right)^n\, I(n),
\ee
where
\be
I(n)=\int\limits_0^1 x^n\left[(2x^2-1)\sqrt{1-x^2}+{{\arcsin}
x\over x}\right]\,dx. \label{v3.28}
\ee
The integration in
(\ref{v3.28}) gives:
\ben &&I(0)={\pi\over 2}\left(\ln 2-{1\over
4}\right),\\ &&I(1)={\pi\over 2}-{16\over 15},\\ &&I(2n)={\pi\over
4n}-{2\Gamma(3/2)\Gamma(n+3/2)\over n \Gamma (n+3)}.
\een
We need the values of $I(n)$ for $n>1$ only for even $n$'s as
starting from $B_3$ all the odd Bernoulli numbers equal zero.
Using the connection of the Bernoulli numbers with Riemann's
$\zeta$-function, after some transformation we find for $\beta^{*}\neq 0$:
\be
\varepsilon&=&-{\hbar c \pi^2\over 16a^4 \beta^{*3}}\Bigg\{
{1\over 2}\left(\ln 2-{1\over 4}\right)-{1\over
\beta^{*}}\left({\pi\over 2}-{16\over 15}\right)\nonumber\\
&+&\sum_{n=1}^\infty (-)^{n-1}{\zeta(2n)\over
\beta^{*2n}2n}\left[1-{(2n+1)!\over
2^{2n-1}(n!)^2(n+1)(n+2)}\right]\Bigg\}\nonumber\\
&=&-{\hbar c \pi^2\over 16a^4 \beta^{*3}}
\left[{1\over 2}\left(\ln 2-{1\over 4}\right)-{1\over \beta^{*}}
\left({\pi\over 2}-{16\over 15}\right)+{\pi^2\over 24\beta^{*2}}+\ldots\right]
\ee

Let us now find the asymptotic expansion of $\varepsilon$ for
small values of the parameter $\beta$ proceeding from formula (\ref{jjj})
written as follows:
\be
\varepsilon&\mathop{=}\limits_{\nu \to 0}&{d^2\over d\nu^2}\left[\sqrt{1+\left({\beta\hbar c\over 2}{d\over d\nu}\right)^2}-
{\beta \hbar c\over2}{d\over d\nu}\right]\nonumber\\
&\times&{1\over \nu}\left({\hbar c\over 4\pi a}+{\hbar c\over 2\pi a}\sum_{n=1}^\infty e^{-\nu\pi n/a}
-{\hbar c\over 2\pi^2} \int\limits_0^\infty dk_z e^{-\nu
k_z}\right).
\ee
Changing the cut-off parameter $\nu'=\nu\pi/a$ after simple calculations reverting further back from
$\nu'$ to $\nu$ we find:
\be
\varepsilon\mathop{=}\limits_{\nu\to 0}{\hbar c \pi^2\over 2a^4}
{d^2\over d\nu^2}\left[\sqrt{1+\left({\beta^{*}}{d\over d\nu}\right)^2}-
{\beta^{*}}{d\over d\nu}\right]{1\over \nu}
\left({1\over 2}+{1\over e^\nu-1}-{1\over \nu}\right).
\ee
Let us make the expansion over small values of the parameter $\beta^{*}$,
applying the procedure from the previous section. We obtain:
\be
\varepsilon={\hbar c\pi^2\over 2a^4}\left(-{1\over 360}+\sum_{k=1}
{\Gamma(k-1/2)(-)^{k-1} B_{2k+4}\over k!\sqrt\pi 2(2k+4)(2k+3)}
\beta^{*2k}\right)
\ee
or, writing explicitly,
\be
\varepsilon=-{\hbar c\pi^2\over 720a^4}\left(1-{1\over
7}\beta^{*2}-{3\over 112}\beta^{*4}-{5\over
264}\beta^{*6}+\ldots\right).
\ee
The first term gives Casimir's result \cite{1},
the remaining terms take into account the field deformation.
As in a one-dimensional space,
the deformation leads to the repulsion and thus diminishes the attraction between the plates.

The numerical calculation of the Casimir energy
\ben
\varepsilon_{\rm 3D}^{*}=\varepsilon\left/{\hbar c\pi^2\over 720
a^4}\right.,
\een
from formula (\ref{v3.24}) is presented in Fig.\,1 (curve 3D).

\section{A two-dimensional case}

Let in the two-dimensional (2D) $xy$ plane the field be concentrated in an infinite
stripe of width $a$ between the straight lines $y=0$ and $y=a$, at
which it equals zero.
From the field transversality only one polarization remains.
It is perpendicular to the wave vector with the components
$k_x,k_y$: $-\infty<k_x<\infty$, $k_y=\pi n/a$, $n=1,2,3,\ldots\;$.
The value $n=0$ is not taken into account, as at $n=0$ the field
equals zero owing to its transversality.

The Casimir energy equals:
\be\label{v4.1}
\varepsilon&=&{1\over 2\pi a}\int\limits_{-\infty}^\infty dk_x \sum_{n=1}^\infty {\hbar c\over 2}\sqrt
{k_x^2+\left({\pi\over a}n\right)^2}\nonumber\\
&\times&
\left[\sqrt{1+\left({\beta\hbar c\over 2}\right)^2
\left(k_x^2+\left({\pi\over a}n\right)^2\right)}
+{\beta\hbar c\over 2}\sqrt{k_x^2+
\left({\pi\over a}n\right)^2}\right]\label{v4.2}\\
&-&{1\over (2\pi)^2}\int\limits_{-\infty}^\infty
dk_x\int\limits_{-\infty}^\infty dk_y
{\hbar c\over 2}\sqrt
{k_x^2+k_y^2}\nonumber\\
&\times& \left[\sqrt{1+\left({\beta\hbar c\over2}\right)^2
\left(k_x^2+k_y^2\right)}+{\beta\hbar c\over 2}\sqrt{k_x^2+k_y^2}\right].\nonumber
\ee
As at $n=0$ the field is absent, the function $f(n)$ must be determined in such a way that $f(0)=0$.
This determination can be made through the cut-off function $e^{-\nu/n}$.
The same determination must be made also in the second term of
Eq.~(\ref{v4.1}). This corresponds to the field energy in a large volume tending to infinity.

Using the Abel--Plana formula after simple transformations similar
to those made in the previous section we obtain from (\ref{v4.2}):
\be
\varepsilon=-{\hbar c \pi\over a^3}\int\limits_0^{1/\beta^{*}}{t^2\, dt\over e^{2\pi t}-1}
\int\limits_0^1 x^2 \sqrt{1-(\beta^{*} t)^2x^2\over 1-x^2}\, dx,
\label{pp}
\ee

At $\beta=0$ the integral over $x$ is easily calculated, it equals $\pi/4$,
and from (\ref{pp}) one obtains the known result \cite{8,10}:
\ben
\varepsilon=-{\hbar c\over 16\pi a^3}\zeta(3).
\een

At $\beta\neq 0$ the integral over $x$ in (\ref{pp})
cannot be expressed in elementary functions. It is reduced to complete elliptic integrals $E(k)$ and $K(k)$
\cite{11,12}.
\be
\int\limits_0^1 x^2 \sqrt{1-(\beta^{*} t)^2x^2\over 1-x^2}\, dx=
{1\over 3}\left(2-{1\over k^2}\right)E(k)+{1\over 3}\left({1\over k^2}-1\right)K(k),
\label{pp1}
\ee
where the modulus $k=\beta^{*}t$.

Let us now find the expansion of $\varepsilon$ at small $\beta$.
For this purpose, one can use the known expansions of elliptic
integrals at small values of the modulus $k$ or proceed directly
from Eq.~(\ref{pp}).
Thus, let us expand in (\ref{pp}) the square root in the numerator
under the integral over $x$ in the series over $\beta^{*}$:
\be
\varepsilon=-{\hbar c \pi\over a^3}\sum_{m=0}^\infty
{\Gamma(3/2)(-\beta^{*2})^m\over
m!\Gamma(3/2-m)}\int\limits_0^1dx\, {x^{2m+2}\over
\sqrt{1-x^2}}\int\limits_0^{1/\beta^{*}}{t^{2m+2}\over e^{2\pi
t}-1}\, dt.
\ee
The first integral over $x$ is equal to
$\Gamma(m+3/2)\Gamma(1/2)/2\Gamma(m+2)$. The second integral over
$t$ can be split into two parts: from 0 to $\infty$ and from $\infty$
to $1/\beta^{*}$, the first one equaling $\Gamma(2m+3)\zeta(2m+3)/2\pi^{2m+3}$ and the
second one gives the terms proportional to $e^{-1/\beta^{*}}$, thus we neglect it at $\beta^{*}\to 0$.
Consequently, after simple transformations we obtain:
\be
\varepsilon&=&-{\hbar c \pi\over a^3}\Bigg[{\zeta(3)\over 16\pi^2}
-\sum_{m=1}{\Gamma(3/2)\Gamma(m-1/2)\Gamma(1/2)\Gamma(m+3/2)\over 2\pi\Gamma(m+2)m!}
\nonumber\\
&\times& {\Gamma(2m+3)\zeta
(2m+3)\over (2\pi)^{2m+3}}\,\beta^{*2m}+O(e^{-1/\beta^{*}})\Bigg]\nonumber\\
&=&-{\hbar c \pi\over a^3}\Bigg[{\zeta(3)\over 16\pi^2}
-{9\zeta(5)\over 128\pi^4}\beta^{*2}-{225\zeta(7)\over
2048\pi^6}\beta^{*4}+\ldots\Bigg].
\ee

Let us proceed to the expansion of the Casimir energy over the
inverse powers of the deformation parameter.
It is simpler to do this directly from formula (\ref{pp}) changing the variable of integration
$y=\beta^{*}t$ and expanding the integrand over the powers of $1/\beta^{*}$ as it was done in the previous cases:
\be
\varepsilon=-{\hbar c \over 2a^3 \beta^{*2}}\sum_{n=0}^\infty
{B_n\over n!}\left({2\pi\over \beta^{*}}\right)^n\, J(n),
\ee
where
\be
J(n)=\int\limits_0^1 \, dx\, {x^2\over \sqrt{1-x^2}}
\int\limits_0^1 \, dy\, y^{n+1}\sqrt{1-x^2y^2}.
\ee
The quantities $J(n)$ for specific values of $n=0,1,2,\ldots$ can be easily calculated:
\ben
&&J(0)={1\over 3}\left({\pi \over 2}-{2\over 3}\right),\\
&&J(1)={{G}\over 4}-{1\over 24},\\ &&J(2)={2\over 15},
\een
here ${G}=0.915965594\ldots$ is Catalan's constant.
Finally at $\beta^{*}\neq 0$ we obtain:
\be
\varepsilon=-{\hbar c \over 2a^3\beta^{*2}}\left[{1\over
3}\left({\pi \over 2}-{2\over 3}\right)- \left({{G}\over
4}-{1\over 24}\right){\pi\over\beta^{*}}+{2\pi^2\over
45\beta^{*2}}+ \ldots\right]. \ee

The results of numerical calculations using formula (\ref{pp}) for
the dimensionless Casimir energy
\ben
\varepsilon_{\rm 2D}^{*}=\varepsilon\Bigg/{\hbar c\over 16\pi a^3}\,\zeta(3),
\een
are presented in Fig.\,1 (curve 2D).
As for the dimensions $D=1$, $D=3$, the Casimir energy $\varepsilon$ tends to zero
from below at $\beta\to \infty$.

\section{Conclusions}

We have shown that the deformation of the field given by Eq.~(\ref{1.2})
leads to the suppression of the Casimir effect for the considered
simple-topology surfaces localizing the field.
The mechanism of this suppression is clear from the fact that
minimal mean-square fluctuations of field strengths equal zero for
the undeformed field and are greater than $\hbar^2 \beta$ for
every vibration mode in the case of the deformed field.
This very fact leads to additional repulsion of the domain
boundaries confining the filed.

The author appreciates interesting discussion with
V.~Tkachuk, Yu.~Krynytskyi, T.~Fityo, and A.~Rovenchak.

\end{document}